\documentclass[aps,prl,preprint,groupedaddress,showpacs,showkeys]{revtex4}
\input epsf

\begin{document}


\title{Shear-strain-induced Spatially Varying Super-lattice Structures on Graphite studied by STM}

\author{S. K. Choudhary and A. K. Gupta}
\affiliation{Department of Physics, Indian Institute of Technology
Kanpur, Kanpur 208016, India.}
\date{\today}

\begin{abstract}
We report on the Scanning Tunneling Microscope (STM) observation
of linear fringes together with spatially varying super-lattice
structures on (0001) graphite (HOPG) surface. The structure,
present in a region of a layer bounded by two straight carbon
fibers, varies from a hexagonal lattice of 6nm periodicity to
nearly a square lattice of 13nm periodicity. It then changes into
a one-dimensional (1-D) fringe-like pattern before relaxing into a
pattern-free region. We attribute this surface structure to
a shear strain giving rise to a spatially varying rotation of the
affected graphite layer relative to the bulk substrate. We propose
a simple method to understand these moir\'{e} patterns by looking
at the fixed and rotated lattices in the Fourier transformed
k-space. Using this approach we can reproduce the spatially
varying 2-D lattice as well as the 1-D fringes by simulation. The
1-D fringes are found to result from a particular spatial dependence of the
rotation angle.

\end{abstract}
\pacs{68.37.Ef, 73.20.At, 73.43.Jn, 61.72.Nn} \keywords{Graphite,
Scanning Tunneling Microscopy, Surface structure, Moir\'{e}
patterns}
\maketitle

\section{Introduction}
Since the discovery of the scanning tunneling microscope (STM) by
Binnig and Rohrer \cite{bin-roh}, highly oriented pyrolytic
graphite (HOPG) has been one of the most commonly used model
surfaces for investigating and modeling of the STM imaging
process. HOPG has a hexagonal structure with weakly coupled layers
of graphene stacked in ABAB... sequence. Graphene is a semi-metal
with zero density of states (DOS) at $E_F$ but a weak inter-layer
interaction in graphite makes this DOS finite but small. This
small DOS is quite sensitive to defects that influence the
overlap of orbitals between layers \cite{kilic}. This makes such defects easily
visible by STM. Further, the weak inter-layer interaction makes
such defects, like moir\'{e} patterns, possible near the surface of HOPG.
A widespread study of this material has led to the discovery of
several types of crystal imperfections native to the basal plane
of HOPG. These include stacking faults, graphite strands and
fibers, broken or flaked layers \cite{chang}, super-lattices
\cite{pong-rev,kuwabara,rong-kuiper,xhie} with periodicity of about
ten nanometers or even larger, and the buckling of the top layer
\cite{pong-buckling}.

Broadly speaking, super-lattice structures in HOPG can occur due
to three reasons. First type appears in graphite-intercalated
compounds \cite{anselmetti} with fixed period super-lattices like,
$\sqrt{3}\! \times \!\sqrt{3}$ or 2$\times$2. Second type is the
electron density oscillations (like Friedal oscillations) near
point defects with a periodicity $\sqrt{3}$ times that of graphite
\cite{mizes,ruffieux}. These oscillations decay rapidly as one
moves away from the defect. The third one involves disorientation
of one or more of the graphite layers near the surface
\cite{kuwabara,rong-kuiper,xhie}. This type of super-lattice
structure is commonly known as moir\'{e} pattern and it arises as
an interference pattern of two identical but slightly rotated
periodic lattices. As first suggested by Kuwabara et. al.
\cite{kuwabara}, this rotation by a small angle $\theta$ gives
rise to a super-lattice of the same symmetry but with much larger
lattice constant given by, $D=d/(2\sin(\theta/2))$, with $d$ as
the lattice constant of the original lattice. This super-lattice
is rotated with respect to the original lattice by an angle
$\phi=30^{\circ}-\theta/2$.

In this paper we present STM study of a spatially varying 2-D
super-lattice pattern and 1-D fringes observed on HOPG (0001)
surface near some defects. Such 2-D patterns have been reported
earlier by several groups and are widely believed to be due to the
moir\'{e} interference \cite{pong-rev,kuwabara,rong-kuiper,xhie}. Spatially varying
periodicity patterns have also been reported before and have been
attributed to a shear strain \cite{bernhardt} or quantum
confinement in a linear potential \cite{harigaya}. The pattern
observed here seems to be a result of an in-plane shear strain
causing a spatially varying moir\'{e} rotation of a top layer.
However, this is the first observation of linear fringes connected
with a moir\'{e} pattern. We also discuss how these fringes can
arise from a shear strain using a simple theory similar to the one
proposed by Amidror and Hersch \cite{amidror1,amidror2} for
moir\'{e} patterns in optics. This approach gives us a better
analytical insight into the large scale structure of such
patterns. An earlier model on moir\'{e} patterns in graphite is
purely computational, based on the variation in local density of
atoms \cite{pong-theory}. In our model, the moir\'{e} patterns are
examined in Fourier transformed k-space to quantify the variation in
local stacking for a fixed as well as a spatially varying
moir\'{e} rotation angle. Using this idea we can understand and
simulate the above observed pattern and, in particular, the 1-D
fringes.

\section{Experimental Details}
Experiments were done with a home built STM similar to the
one described elsewhere \cite{gupta-ng}. This STM uses a
commercial electronics and software \cite{rhk}. The data reported
here were taken in ambient conditions. HOPG was fixed on the
sample holder with a conducting epoxy and the sample was freshly
cleaved using an adhesive tape before mounting it on the STM. Fresh
cut Pt$_{0.8}$Ir$_{0.2}$ wire of 0.25 mm diameter was used as the
STM tip. The images have been obtained in constant current
(feedback on) mode. The images shown here are filtered to remove
steps and spikes, however, for quantitative analysis we have used
the unfiltered data.

\section {Results}

\subsection{Spatially varying lattice and 1-D fringes}

\begin{figure}
\centerline{\epsfxsize = 2.5 in \epsfbox{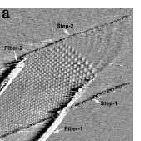} \hspace{0.4cm}
\epsfxsize = 2.5 in \epsfbox{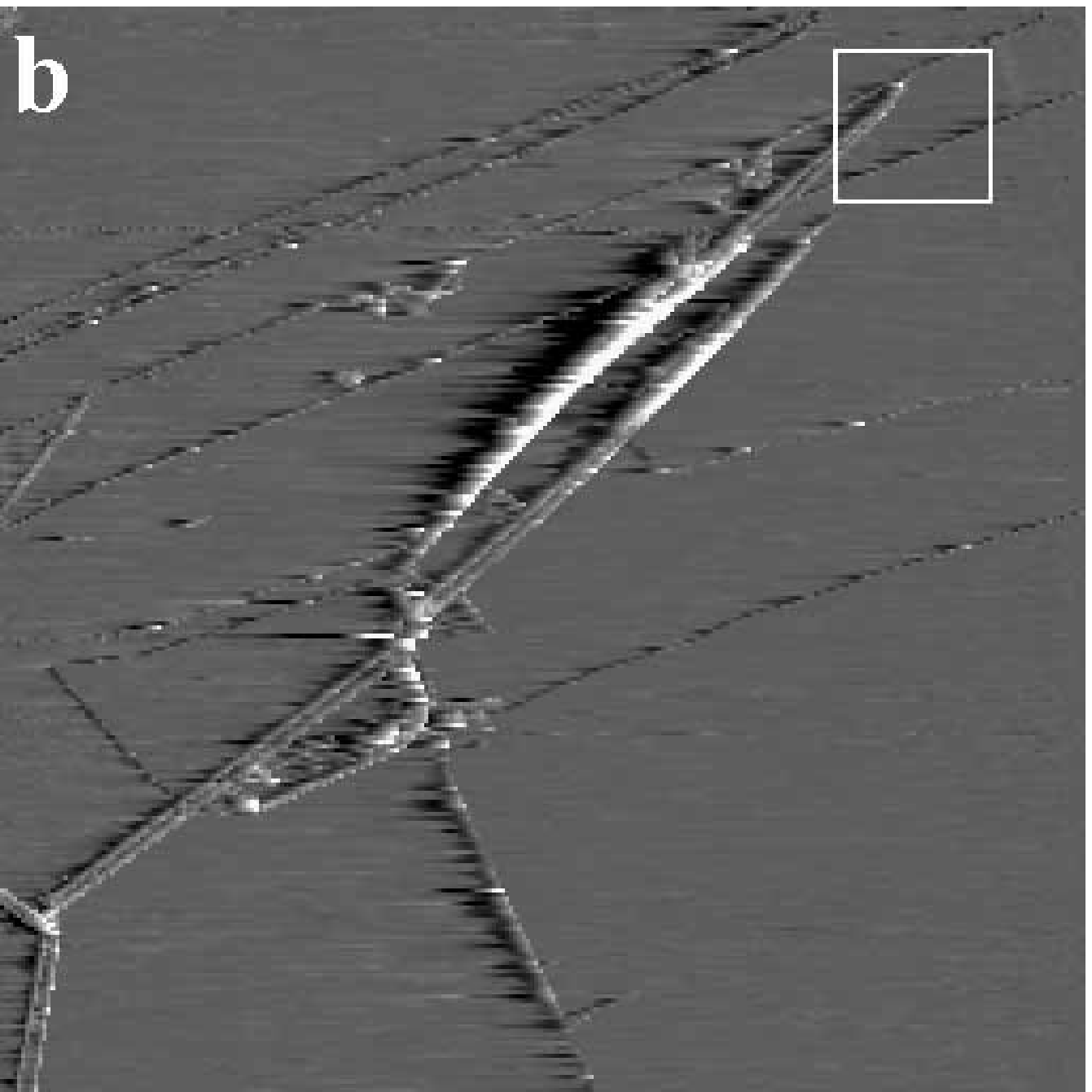}}
\caption{\label{fig:suplatt} {\bf a.} A STM image
(0.32$\times$0.32 $\mu m^2$ at 0.5V and 0.2nA) showing a spatially
varying super-lattice structure confined between fiber-1 and
fiber-2 on the layer defined by step-1 and step-2. Near the
termination of superstructure a 1-D fringe pattern appears. {\bf
b.} A later STM image (2.16$\times$2.16 $\mu m^2$ at 0.8V and
0.2nA) showing a larger portion of fiber-1 and fiber-2. The area
of left image marked by the white square.}
\end{figure}

Fig.\ref{fig:suplatt}a shows a topographic image of (0001) basal
plane of HOPG with the super-lattice structure together with two
steps and two carbon fibers. We mark these as fiber-1 \& 2 and
step-1 \& 2. The step-1 is a double layer step with a height of
0.7$\pm$0.1nm while the step-2 is a single layer step. These step
heights have been found from the images taken away from the
super-lattice pattern area. These two steps divide the area into three terraces with
the middle one having a super-lattice pattern confined between the
two fibers as seen in the same figure.

Carbon fibers have been observed earlier by various groups on the
graphite surface \cite{pong-rev}. These, we believe, are long and
thin ribbons or rolls of graphite created, presumably, during the
cleaving process. The exact vertical location of the fibers
relative to various visible layers is not so clear in this image.
However, since the fibers are strictly limiting the pattern on the
mono-layer terrace defined by step-1 and step-2, we believe that
these fibers are in contact with this affected layer. They may
be either below or above this particular layer. If a layer is
above the fiber it would go over the fiber with some buckling.
Such buckling cannot be ruled out by stress considerations as the
fibers are quite wide ($\ge$10nm) and not so high ($\le$1nm).

A relatively larger area image including the patterned region is shown in Fig.\ref{fig:suplatt}b.
This image was taken later when the fiber-1 was fully relaxed as
discussed in detail later. The region of the left image is marked
by a square in the right one. The other ends of the two fibers are
also visible in Fig.\ref{fig:suplatt}b. Here we see that the
fiber-2 looks brighter above the step-1 and the fiber-1 has
similar behavior but this is seen in the Fig.\ref{fig:suplatt}a.
Here, both the fibers look less bright when they emerge from
the covered layer(s) near the center of the image in
Fig.\ref{fig:suplatt}b.

As shown in Fig.\ref{fig:suplatt}a the super-lattice contains two
regions namely a 2-D lattice extending till the end of fiber-1 and
a 1-D wave-like pattern starting from the fiber-1 end. The 1-D
fringes bend towards the step-2 and terminate at this step. The
2-D lattice continues beyond the fiber-2 end but terminates at
step-2 in this image. The 1-D pattern contains the same number of
maxima and minima as the terminating 2-D lattice. The 2-D lattice
is not uniform as it evolves from a hexagonal lattice deep inside
the fibers to nearly a square lattice with much larger
periodicity. The super-lattice structure is also present over the
step-1 (up to fiber-1) but with corrugation reduced by a factor of
2.3. Although this is not clearly visible in Fig.\ref{fig:suplatt}
but we clearly see it on zooming into this area. The corrugation
is also found to vary with periodicity as analyzed later.

\subsection {Pattern variation with fiber-1 motion}

\begin{figure}
\epsfxsize = 1.8in \epsfbox{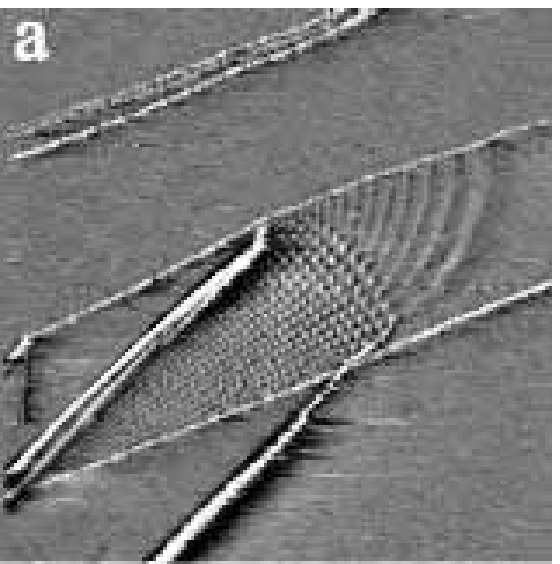} \hspace{0.2cm} \epsfxsize =
1.8in \epsfbox{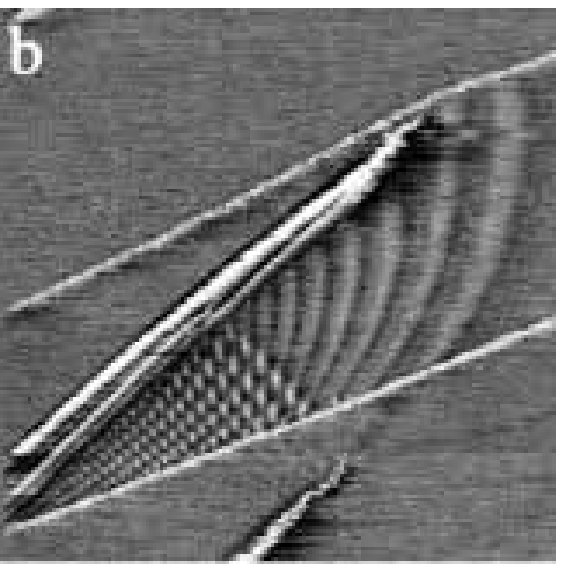} \hspace{0.2cm} \epsfxsize = 1.8in
\epsfbox{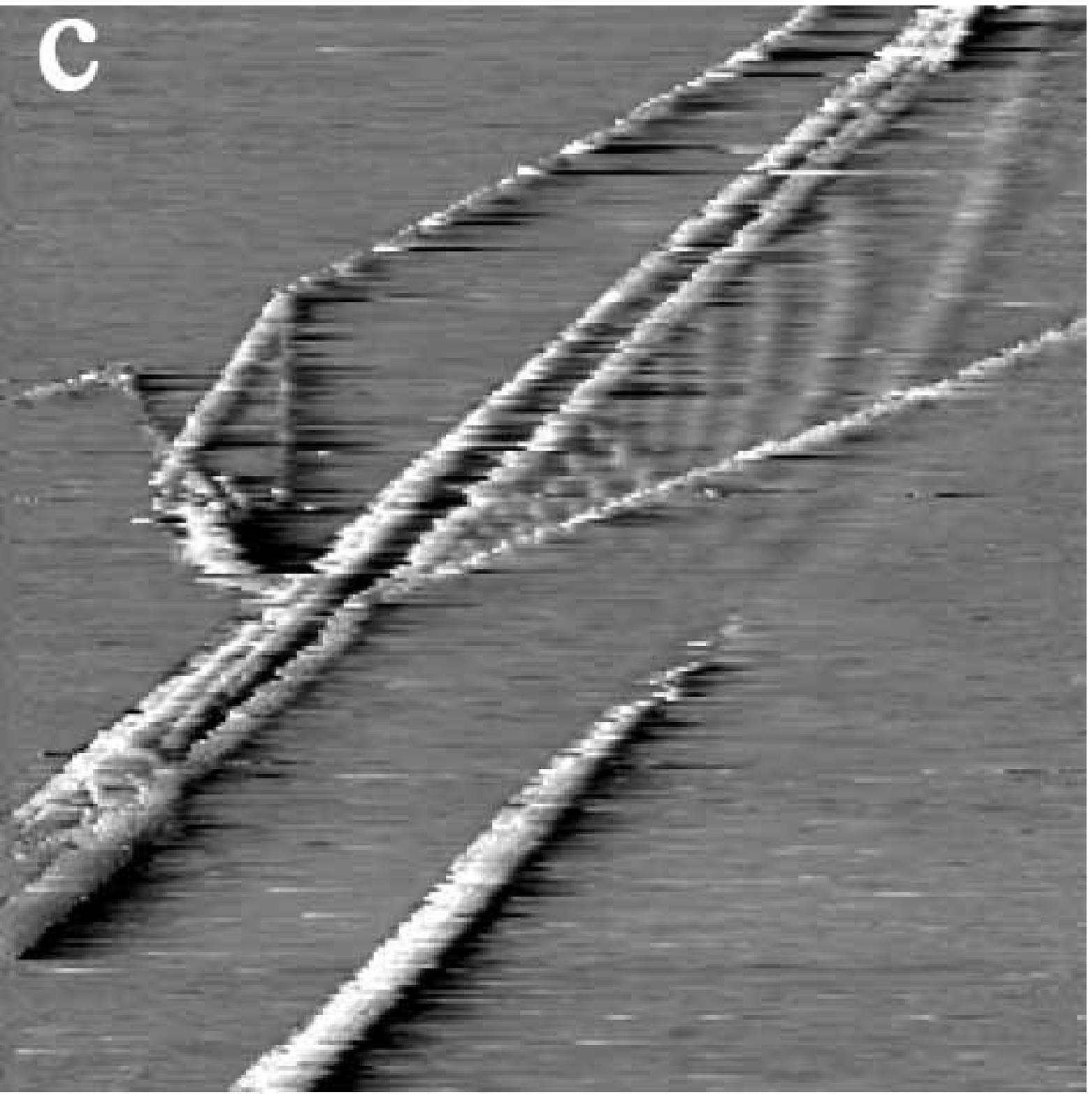} \caption{\label{fig:etching} The topographic
images of the super-lattice structure showing its evolution with
the motion of fiber-1. {\bf a.} 0.52$\times$0.52$\mu m^2$, {\bf
b.} 0.35$\times$0.35$\mu m^2$, {\bf c.} 0.43$\times$0.43$\mu m^2$
taken at 0.5V and 0.2nA.}
\end{figure}

As we image this area repeatedly, fiber-1 becomes shorter in
length as the whole fiber seems to recede. This may be due to some
stress on the fiber. Several groups have reported various surface
modifications in the STM experiments but the exact mechanism
behind such modifications is not quite understood. We show three
of the latter images in Fig.\ref{fig:etching} depicting the
changing pattern with the fiber-1 withdrawal. The withdrawal of
this fiber is not uniform with time as sometimes a significant
length disappears in one scan and at other times it recedes
gradually. Eventually the receding stopped and the length of the
fiber stayed same over a couple of days. One remarkable
observation is the pinning of the boundary between the 2-D
super-lattice and the 1-D fringes to the fiber-1 end. With the
receding of fiber-1, the 1-D fringes terminate on the fiber-2 as
opposed to step-2 in earlier images. Moreover, the spatial extent
and corrugation amplitude of 1-D fringes increases together with
some modifications in the large scale structure of the 1-D
pattern.

\section {Analysis}
\begin{figure}[htbp]
\centerline{\epsfysize=1.8in \epsfbox{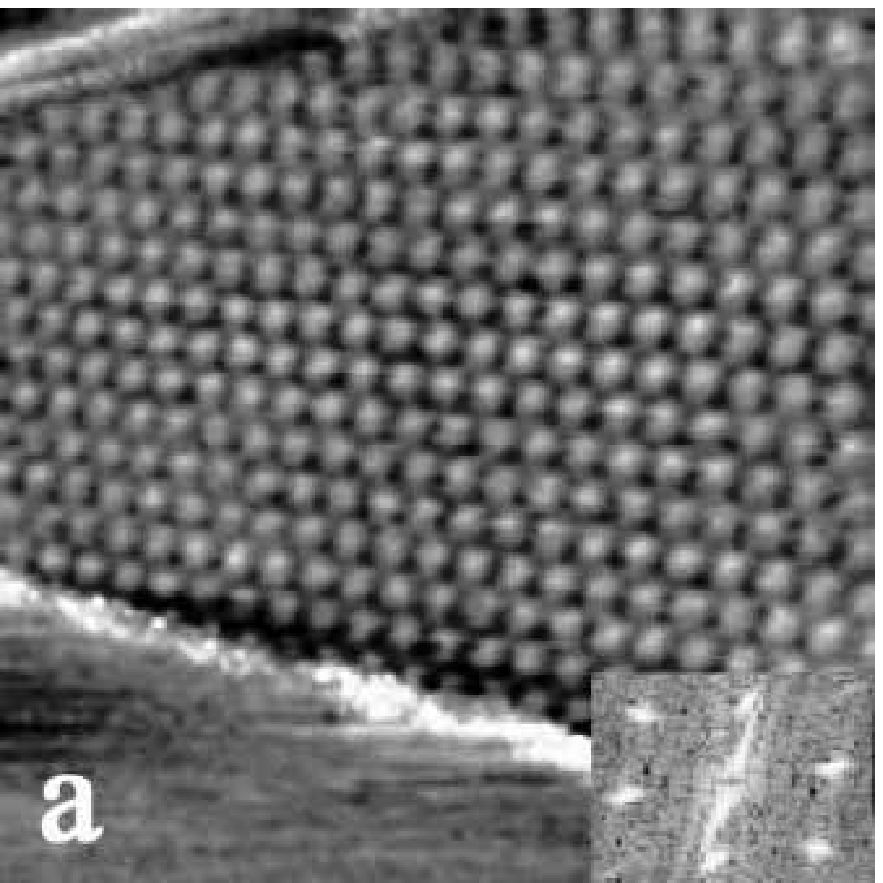} \hspace{0.07cm}
\epsfysize=1.8in \epsfbox{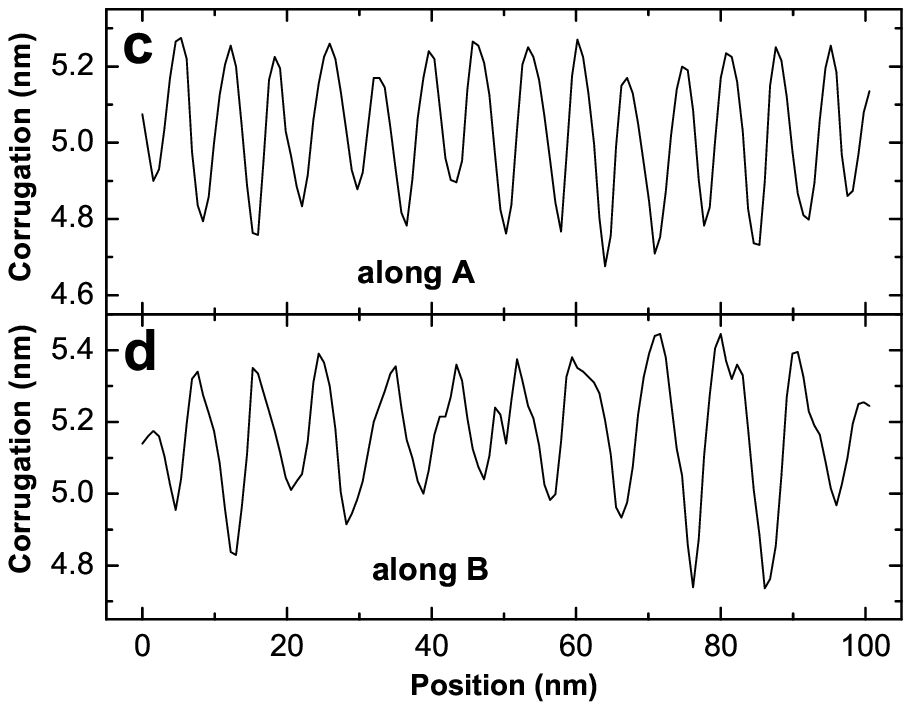}}\vspace{0.2cm}
\centerline{\epsfysize=1.8in \epsfbox{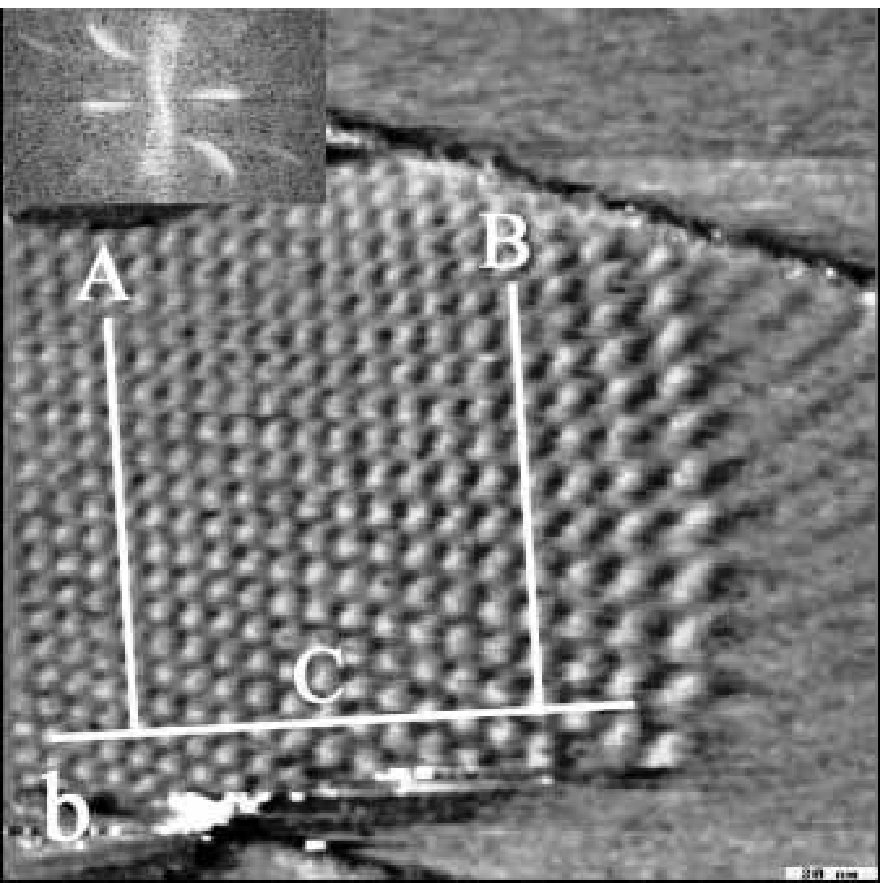}
\hspace{0.28cm}\epsfysize=1.7in \epsfbox{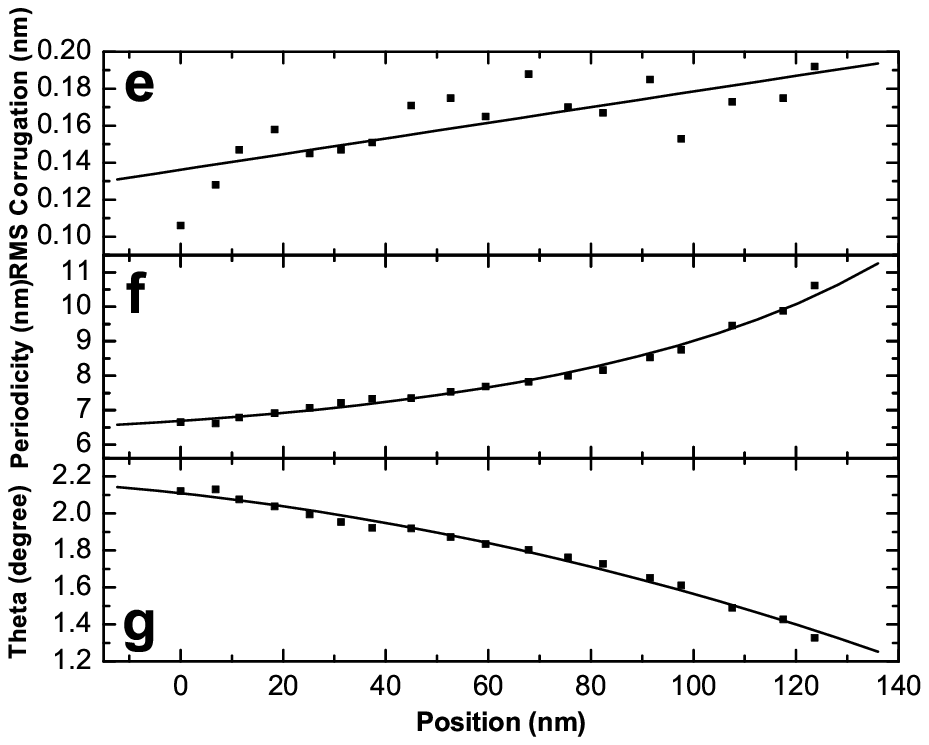}}
\caption{\label{fig:fft}{\bf a.} Super-lattice pattern image
(120$\times$120nm$^2$ at 0.5V and 0.2nA) taken deep inside the
fibers, and {\bf b.} image (196$\times$196nm$^2$, 0.5V/0.2nA)
taken near the boundary of the pattern. The insets show the
respective Fourier transforms. {\bf c} and {\bf d} show the line
scans along the lines A and B. {\bf e}, {\bf f} and {\bf g} show
the variation (along line C) in rms corrugation, periodicity, and
rotation, respectively.}
\end{figure}

Two close-up images of this pattern are shown in Fig.\ref{fig:fft}
in two different regions of the structure at different orientation
of the image window as compared to the previous ones. These images
have been taken at an early stage when fiber-1 was extending the
most. As apparent from the image shown in Fig.\ref{fig:fft}a, the
pattern deep inside the fibers (towards left) is a hexagonal
lattice of fixed periodicity. As we move out, the periodicity
increases with the pattern evolving to an oblique lattice and then
to nearly a square lattice as seen in Fig.\ref{fig:fft}b. Fourier
transforms (FT) of these images as shown in the inset also
illustrate this variation. In detail, the FT in Fig.\ref{fig:fft}a
shows six sharp spots while the FT in Fig.\ref{fig:fft}b has these
spots elongated.

To analyze the periodicity variation carefully, line cuts were
taken along various lines perpendicular to the line `C' in
Fig.\ref{fig:fft}b. Two such representative lines are A and B
along which the topographic height is plotted in
Fig.\ref{fig:fft}c and Fig.\ref{fig:fft}d. These lines were
carefully chosen to pass over the bright spots of the 2-D lattice.
The periodicity along the lines
perpendicular to C was found to be visibly constant but as one
moves out along C this periodicity increases. The corrugation
along the line C also varies. The RMS corrugation and periodicity
as measured by line cuts perpendicular to C are plotted in
Fig.\ref{fig:fft}e and Fig.\ref{fig:fft}f as a function of
distance along the line C. The periodicity along such lines has
been found from the average separation between peaks.

If we use the moir\'{e} rotation hypothesis, the rotation angle
$\theta$ (between the top layer and the layers underneath) should
also change spatially to get a varying lattice spacing D. Using
$D=d/(2\sin(\theta/2))$ (d=0.246nm), we plot the variation of
$\theta$ in Fig.\ref{fig:fft}g. This varying $\theta$ implies that
this rotated layer undergoes a shear as we move out of the fibers.
The portion of this layer that is well inside the fibers has a
constant rotation and the portion far outside the fibers is free
of any moir\'{e} pattern and thus $\theta$ is zero. The
intermediate region is strained with a changing $\theta$ and
periodicity. As estimated from the periodicity, the angle $\theta$
changes from $2.3^\circ$ to $1.3^\circ$. The 1-D fringe pattern
which has a close correlation with the 2-D moir\'{e} pattern
prompts us towards a common origin for the two patterns. We
believe that $\theta$'s particular spatial dependence is
responsible for the 1-D fringes as discussed later.

A second order polynomial fit (as shown in Fig.\ref{fig:fft}g) for
this $\theta$ variation gives an expression,
\begin{eqnarray}
\theta=A_0-A_1x-A_2x^2 \label{eq:poly}
\end{eqnarray}
with $A_0=(2.11\pm0.02)$degree,
$A_1=(3.0\pm0.6)\times10^{-3}$degree/nm,
$A_2=(2.4\pm0.5)\times10^{-5}$degree/nm$^{2}$. This is also the
expression for local angular orientation for a cantilever with one
end clamped and the other end loaded \cite{cantilever}.

A possibility on how the stress arises in the
graphite layer is similar to the bending of a cantilever
with certain loading conditions caused by the fibers. By  the definition of strain, the maximum strain at a
given $x$ would be $\frac{d}{2}\frac{d\theta}{dx}$. To estimate the maximum strain, the $\frac{d\theta}{dx}$ is maximum at the boundary of the 2-D super-lattice and 1-D fringes and it is found from the polynomial fit
(Eq.\ref{eq:poly}) to be (1.6$\pm$0.3)$\times$10$^{-4}$rad/nm. At
this location the thickness (d) is 115nm. This point gives the largest stress,
$\sigma_{max}$=1.1$\pm$0.2 GPa, using $Y$=121.9 GPa
\cite{synder-prb} for graphite. For comparison, the
macroscopic indentation experiments indicate that stress on the
order of $\approx$1 GPa damages the surface of graphite
\cite{skinner}. On the other hand, Snyder et. al.
\cite{synder-prb} have shown that the stress required to induce
dislocation motion on basal plane of HOPG is of order 5-200 MPa.

\section{Understanding the linear fringes from spatially varying moir\'{e} rotation}
\begin{figure}
\centerline{\epsfxsize=1.5in \epsfbox{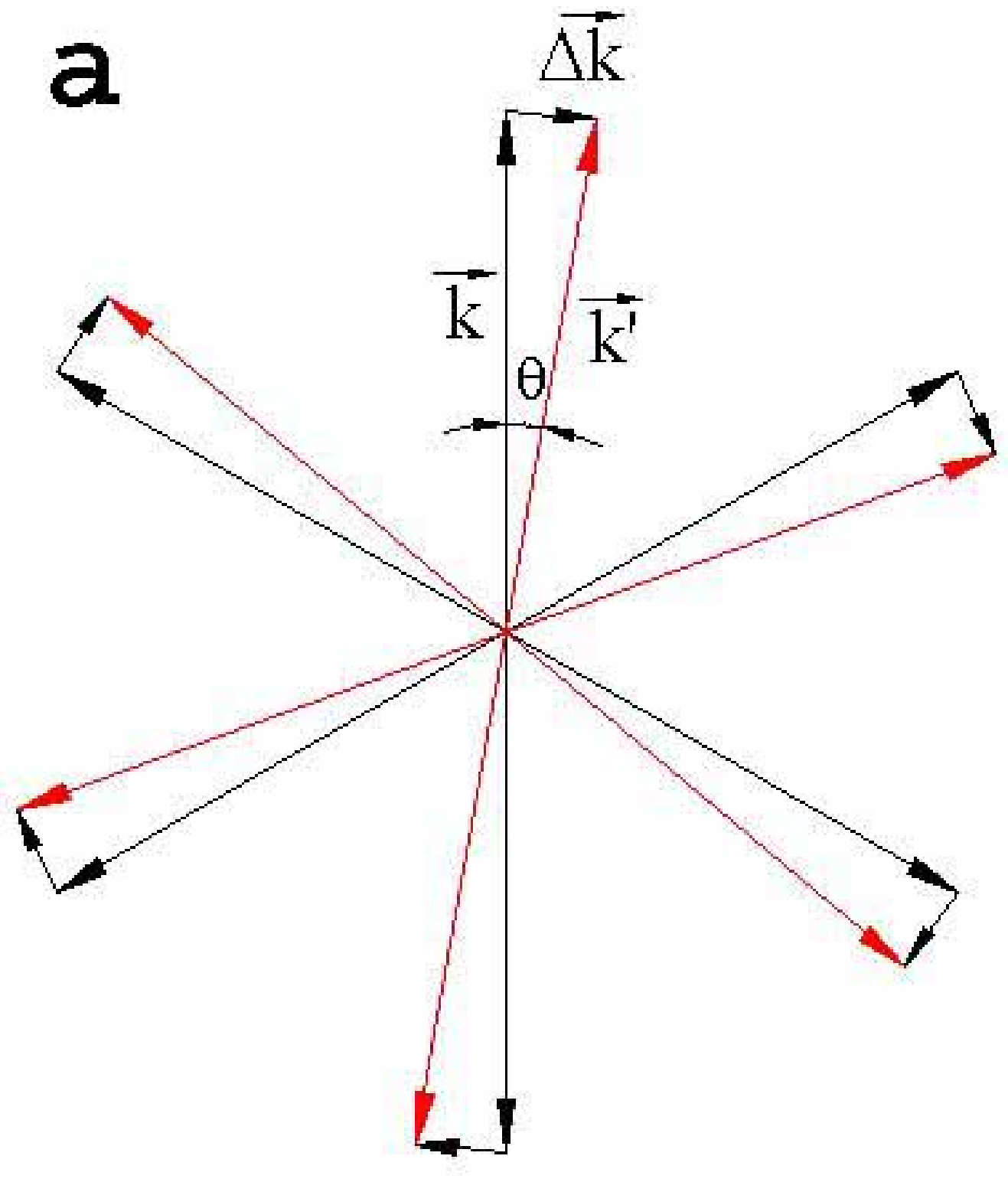} \hspace{0.4cm}
\epsfxsize=1.5in \epsfbox{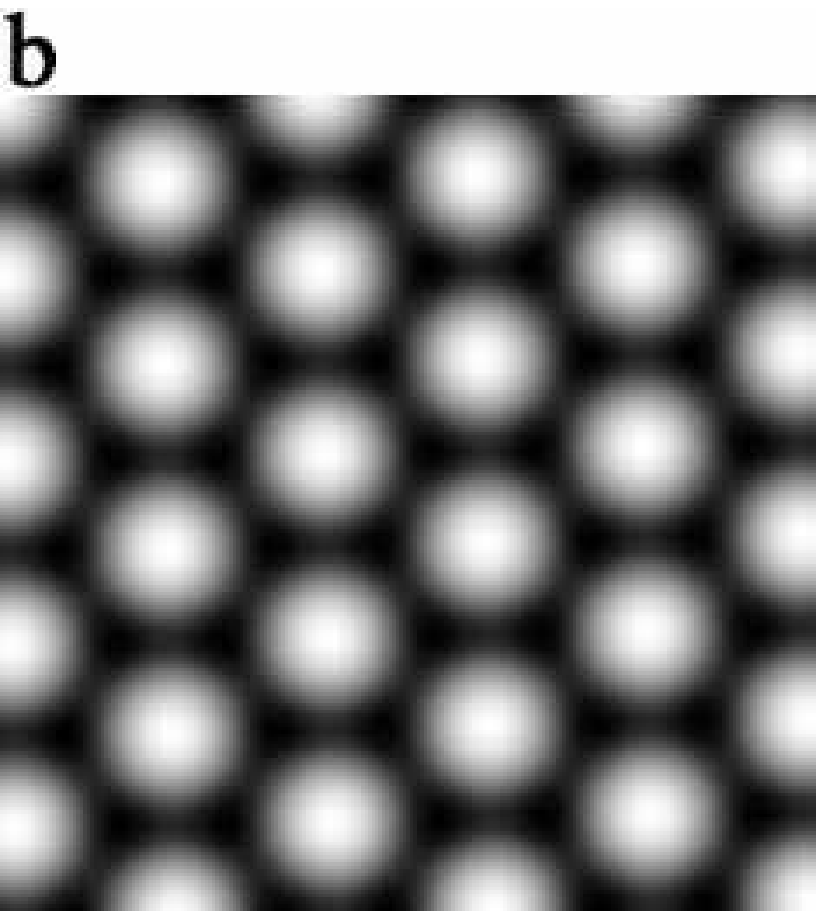}} \caption{\label{fig:moire}
{\bf a} The lattice vectors of the fixed ($\bf k$) and slightly
rotated ($\bf k'$) hexagonal lattices and their differences ($\bf
\Delta k$). In this figure y-axis is pointing vertically up and x
is pointing to the right. {\bf b} Moir\'{e} interference pattern
calculated using Eq.\ref{eq:moire-fft} with a
$\theta$=2.5$^\circ$. The size of the image is
25$\times$25nm$^2$.}
\end{figure}

Here we describe a simple model to understand and to simulate the
large-scale structure of the spatially varying moir\'{e} patterns.
For illustration, let us first consider a 1-D periodic lattice. If
we superimpose two 1-D periodic patterns given by $\cos k_{1}x$
and $\cos k_{2}x$ ($\mid k_{1}-k_{2}\mid\ll k_1,k_2$), we would
see an interference (or beats-like) pattern with periodicity
$2\pi/|k_{1}-k_{2}|$ in the superposed pattern (i.e., $\cos
k_{1}x\cos k_{2}x$). In 2-D if we superimpose two hexagonal
patterns of slightly different reciprocal lattice vectors ${\bf
k_{i}}$ and ${\bf k_{i}'}$ (i = 1-6), the resulting pattern in
real space would come from the difference of the two sets of
lattice vectors, i.e., (${\bf k_{i}'-k_{j}}$). The longest period
variations would arise from $\Delta {\bf k_{i}}= {\bf
k_{i}'-k_{i}}$ (see Fig.\ref{fig:moire}), which are actually responsible for the moir\`{e}
patterns here. In 2-D, the pattern could arise due to
differences in both the magnitude and direction of ${\bf k}$ and
${\bf k'}$ \cite{amidror1} but in the present case only the
direction mismatch is playing the major role.
$\Delta {\bf k_{i}}$ as seen from Fig.\ref{fig:moire}, has a
magnitude of $2k\sin (\theta/2)$ and the smallest angle it makes
with one of the ${\bf k_{i}}$'s is 30$^{\circ}\! -\theta$/2. This
is consistent with the aforementioned periodicity and angle of the
moir\'{e} super-lattice. Now the moir\'{e} interference pattern in
real space is easily produced by inverse Fourier transformation
(IFT) of these six ${\Delta {\bf k_i}}$ vectors. This IFT gives a
pattern described by,
\begin{eqnarray}
I(x,y)=I_{0}\,+I_1[\cos(f{_1})+\cos(f{_2})+\cos(f{_3})],
\label{eq:moire-fft}
\end{eqnarray}
where $I_0$ and $I_1$ are two constants and
\begin{eqnarray}
f{_n}=2k\sin(\theta/2)\{-y\sin((n-1)\pi/3\!+\!\theta/2)\!+\!x\cos((n-1)\pi/3\!+\!\theta/2)\}].
\label{eq:moire-f-def}
\end{eqnarray}
Here n=1, 2 or 3 and $k=4\pi/\sqrt{3}d$ with $d$ (=0.246nm) as the
real space ab-plane lattice parameter of graphite. Each cosine
term in Eq.\ref{eq:moire-fft} gives rise to a 1-D periodic pattern
with bright and dark fringes along parallel straight lines (for
constant $\theta$). The bright fringes of $\cos f_n$ are described
by the contours $f_n(x,y)=2N\pi$, with N as an integer. For a
constant $\theta$, the fringes due to the three cosine terms make
an angle $\pi/3$ with each other and the sum of
these three cosines leads to a 2-D triangular lattice pattern. One
such pattern is shown in Fig.\ref{fig:moire}b for
$\theta=2.5^\circ$, $I_0$=0, and $I_1$=1. In this model, the
$I(x,y)$ only quantifies the local stacking pattern. The $I(x,y)$ has a variation between +3 and -3
(taking $I_0=0$ and $I_1=1$) with a value of +3 signifying an AA
stacking and -3 signifying an AB stacking. An intermediate
value represents a slip with the degree of slip quantified by the
magnitude of $I(x,y)$. The DOS at the surface is a result of top
three layers and so the DOS at BAB and BAC could be actually
different while our model does not differentiate between these two
stacking sequences \cite{rong-kuiper,xhie}.

This model can also be used for the geometrically transformed
lattices \cite{amidror2}, where the rotation angle is spatially
varying. This way we can model the spatially varying moir\'{e}
patterns and understand their large scale structure. In
particular, we can find an analytical condition on the spatial
variation of the rotation angle that gives rise to the 1-D fringes
as following. For spatially varying $\theta$, the 1-D fringes due
to each $\cos f_n$ will have a spatially varying periodicity and,
in general, these fringes may not be straight and parallel to each
other. These fringes will be locally perpendicular to $\nabla f_n$
and their local periodicity can be quantified by $2\pi/|\nabla
f_n|$. These $f_n$'s (see Eq.\ref{eq:moire-f-def}) satisfy the
condition,
\begin{eqnarray}
f_1=f_2-f_3. \label{eq:grad-f}
\end{eqnarray}
For $I(x,y)$ to have a 1-D pattern, the fringes due to the contributing cosine terms must be locally parallel to each other or, in other words, each $f_n$ should have the same constant value contours. In particular, and as it turns out to be the case for the observed 1-D fringes, if the $\theta$ variation is such that $f_1=C$ (a constant) then using Eq.\ref{eq:grad-f}, $f_2=C+f_3$. For such $\theta$ variations $\cos f_1$ gives no fringes while the fringes due to $\cos f_2$ and $\cos f_3$ are locally parallel with the same local wavelength. Thus $I(x,y)$ would have only 1-D fringes. Similar argument would also hold for $f_2$=constant or $f_3$=constant, which is obvious from symmetry considerations. The expression for $f_1$ (Eq.\ref{eq:moire-f-def}) can be simplified to,
\begin{eqnarray}
f_1=kx\sin\theta+ky(\cos\theta-1), \label{eq:f1-simplified}
\end{eqnarray}
and so for $f_1(x,y)=C$ we find,
\begin{eqnarray}
\theta(x,y)=sin^{-1}\left[\frac{(C/k)+y}{\sqrt{x^2+y^2}}\right]-tan^{-1}\left(\frac{y}{x}\right).
\label{eq:theta-lin-fr}
\end{eqnarray}
In small $\theta$ approximation we can neglect the ($\cos\theta-1$) term in comparison to $\sin\theta$ term in Eq.\ref{eq:f1-simplified} and this gives $\theta(x,y)=C/kx$. Here, one has to be careful in choosing the origin while using this form of $\theta(x,y)$. The same origin has to be used in calculating the $I(x,y)$ to ensure 1-D fringes. In fact this choice of origin gives one more parameter in choosing $\theta(x,y)$.

\begin{figure}[htbp]
\centerline{\epsfysize=1.5in \epsfbox{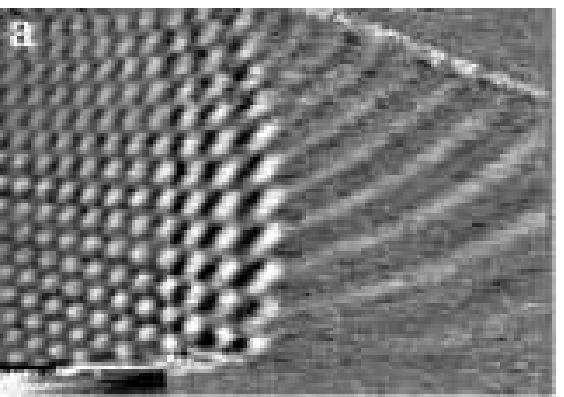} \epsfysize=1.5in
\epsfbox{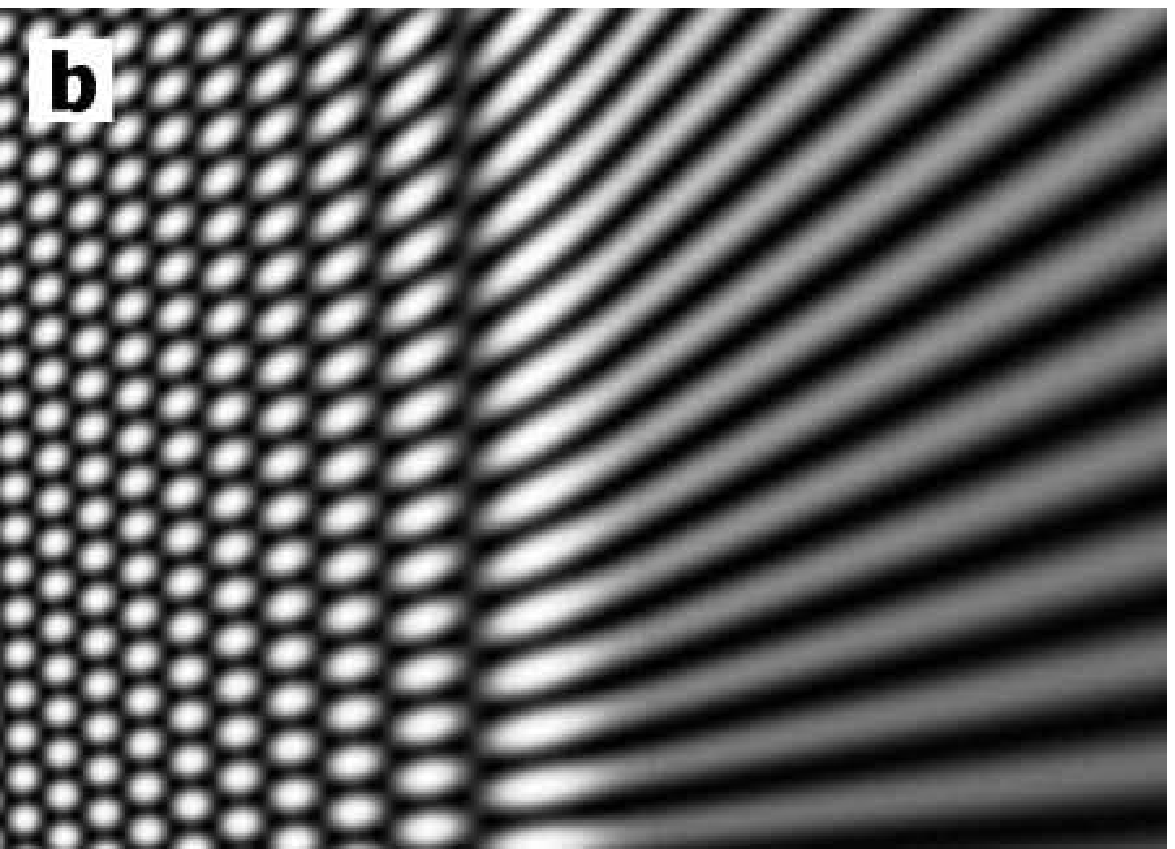} \epsfysize=1.5in
\epsfbox{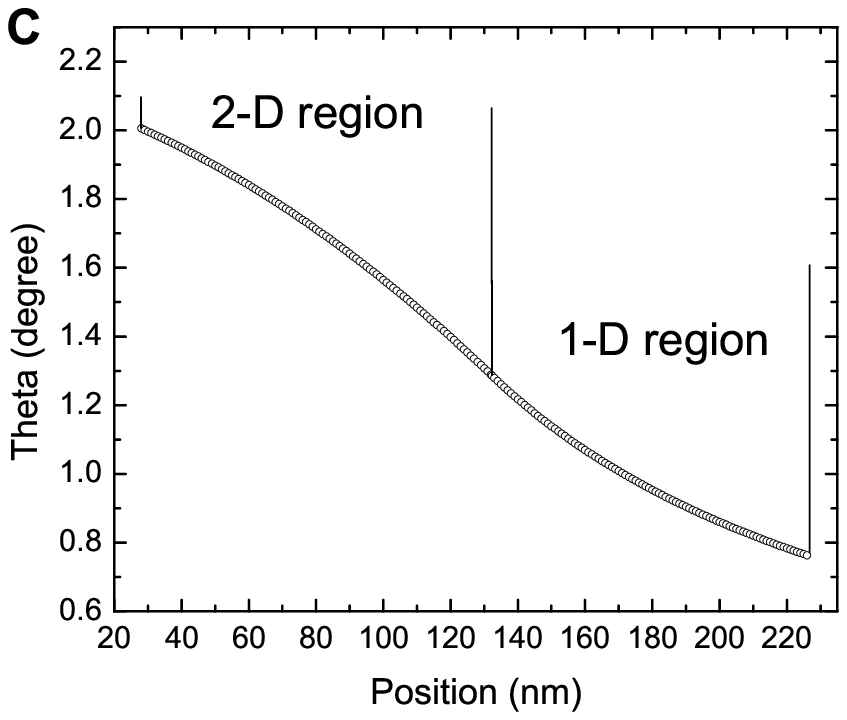}}\vspace{0.2cm} \centerline{\epsfysize=1.5in
\epsfbox{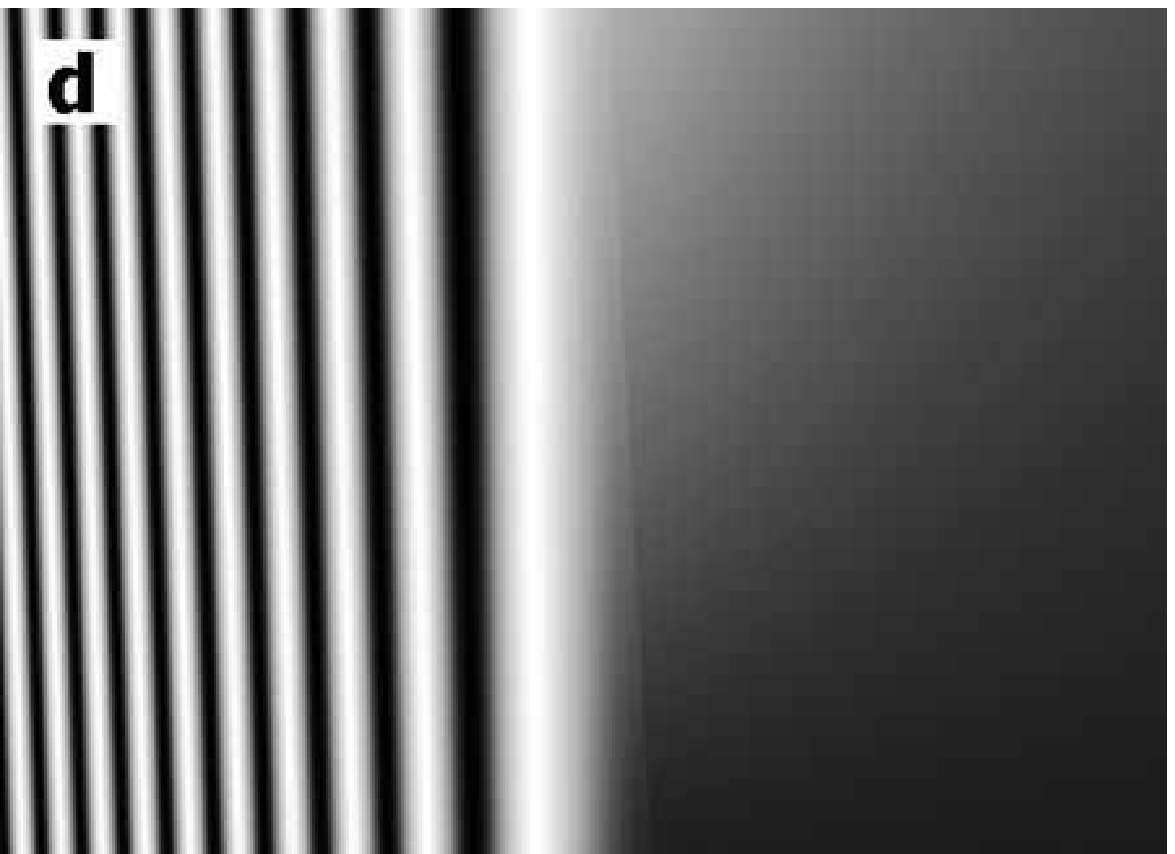} \epsfysize=1.5in \epsfbox{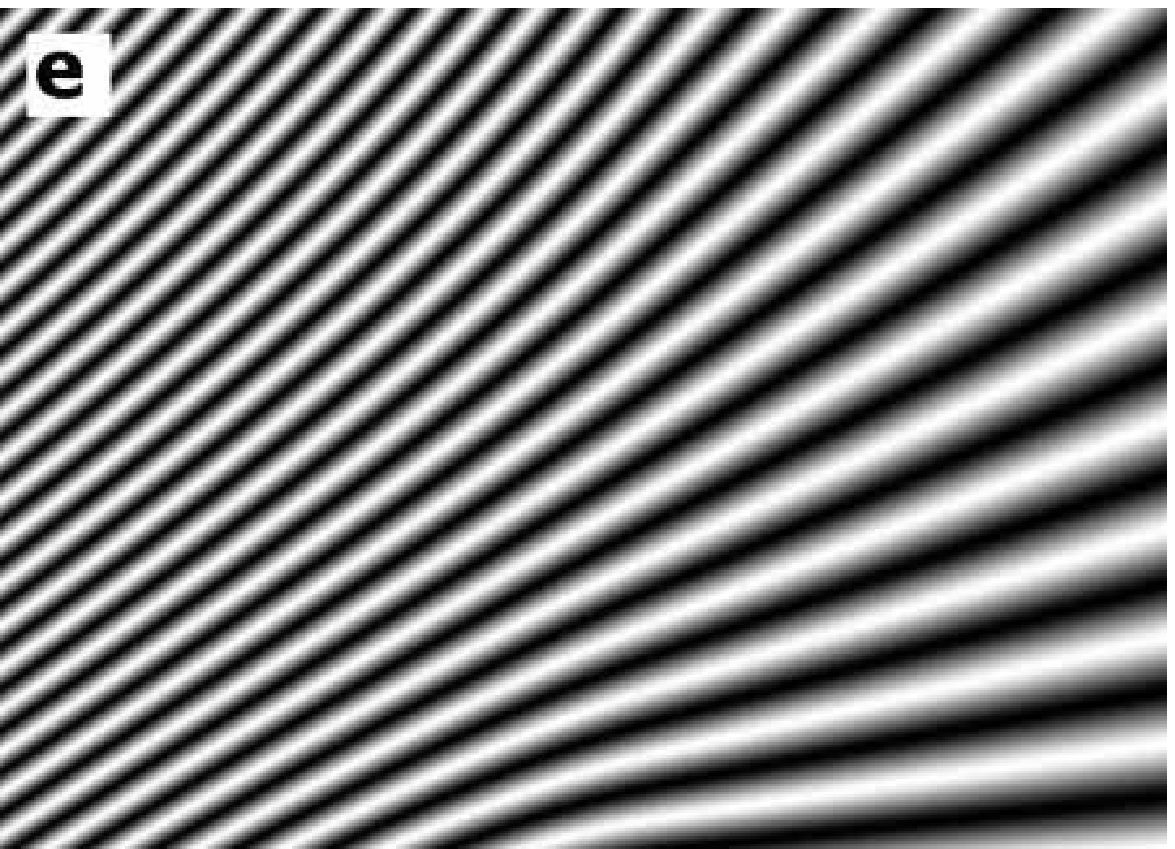}
\epsfysize=1.5in \epsfbox{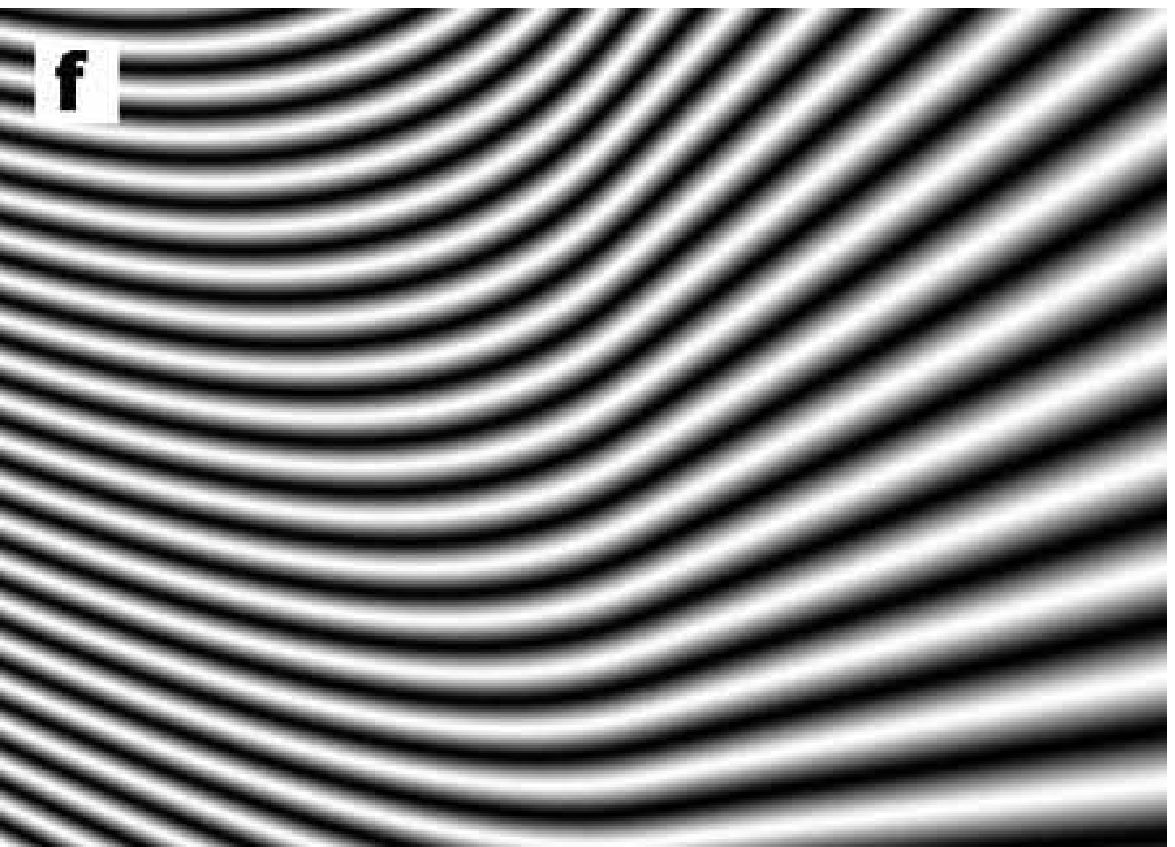}} \caption{\label{fig:simul}
{\bf a.} STM image of the super-lattice pattern of varying periodicity and linear fringes (196$\times$139nm$^2$ at 0.5V and 0.2nA). The 2-D pattern is present up to 100 nm and beyond this it changes into linear fringes {\bf b.} Simulated image (see text) of size 198$\times$141nm$^2$ with the $\theta$ variation shown in {\bf c}. {\bf d, e, f} show the three cosine components, i.e., $\cos f_1$, $\cos f_2$ and $\cos f_3$, respectively, of the image in {\bf b.} (In figures {\bf b, d, e} and {\bf f} y-axis is pointing at angle 3$^\circ$ to vertically up and x-axis is perpendicular to y and towards the right.}
\end{figure}

A simulated pattern using above ideas is shown in Fig.\ref{fig:simul}b together with the observed pattern in Fig.\ref{fig:simul}a. Here, we have used two different spatial forms of $\theta$ in different x intervals. We have used $\theta(x,y)$ as found in Eq.\ref{eq:poly} for 2-D region, where $\theta$ changes from 2$^\circ$ to 1.29$^\circ$. We use the small $\theta$ approximation for 1-D fringes, i.e., $\theta(x,y)=C/k(x+a)$ for $\theta$ between 1.29$^\circ$ and 0.76$^\circ$. The constants $C/k$ and $a$ are found to be 3.067 rad-nm and 4.399 nm, respectively, by ensuring the continuity of $\theta$ and $d\theta/dx$ at x=132.3 nm. This is necessary to keep the stress finite and continuous everywhere. For 1-D fringes  the required $\theta(x,y)$ is of the form of $C/kx$, so in Fig.\ref{fig:simul}, we have actually shown images of $I(x+a,y)$ with $a=4.399$ nm.  A plot of the used $\theta$ variation as a function of $x$ is shown in Fig.\ref{fig:simul}c. Since the 2-D to 1-D pattern boundary is pinned to the fiber-1 end we believe that there is a sudden change in spatial dependence of stress across this $x$ causing a change in $\theta$ behavior that in turn is responsible for the boundary. Fig.\ref{fig:simul} also shows the three $\cos f_n$ maps, where one can see clearly how the pattern arises in the two regions in terms of the three cosines as argued before. In the 1-D fringe region, we see that $\cos f_1$ gives no fringes while $\cos f_2$ and $\cos f_3$ have fringes that are locally parallel to each other. In the 2-D pattern region the fringes due to the three cosines are not parallel.

So far we discussed a particular condition on $\theta$ that gives 1-D fringes. Now we find a more general condition on $\theta$ for getting 1-D fringes and we also argue here that above used spatial dependence of $\theta$ for 1-D fringes is the only possibility (in small $\theta$ approximation) for the observed fringes. The existence of 1-D fringes requires that all the constant value contours of $f_1$, $f_2$, and $f_3$ are identical. This is possible if $f_1= \phi(f_3)$ and, using Eq.\ref{eq:grad-f}, $f_2= f_3+\phi(f_3)$, where $\phi$ is a scalar function. If we assume $\phi$ to be a linear function, i.e., $\phi(x)=\alpha(x+C_1)$ then $f_1-\alpha f_3$=$\alpha C_1$. This, using Eq.\ref{eq:moire-f-def}, gives the same solution for $\theta(x,y)$ as Eq.\ref{eq:theta-lin-fr} except for $x$, $y$, and $C$ replaced by $x'=x(\alpha^{-1}+1/2)-y(\sqrt{3}/2)$, $y'=y(\alpha^{-1}+1/2)+x(\sqrt{3}/2)$ and $C_1$, respectively. In small $\theta$ approximation, this gives $\theta(x,y)=C_1/kx'$. Therefore, any function of the form $\theta(x,y)\propto1/(x+\gamma y)$ would give rise to linear fringes for arbitrary $\gamma$ (=$\sqrt{3}\alpha/(\alpha+2)$). The form of $\theta$ used in producing the linear fringes in Fig.\ref{fig:simul} is a special case with $\alpha$=0 and $C_1 \rightarrow \infty$ with $\alpha C_1$=C. From Fig.\ref{fig:simul}a we see that the observed 1-D fringes connect smoothly to the rows of bright spots coming from $\cos f_2$ and $\cos f_3$ while the rows of bright spots due to $\cos f_1$ terminate before the 1-D fringes. This means that $\cos f_1$ is not contributing to the observed 1-D fringes and $\cos f_1$ has to be constant in this region. So $\theta=C/kx$ is the only possibility for the observed fringes in small $\theta$ approximation.

\section{Discussion}
Our simulation nicely captures most of the details of the observed pattern except for the curvature of the 1-D fringes. For the curvature in 1-D fringes, we find that any modification in spatial dependence of $\theta$ from $C/kx$ affects the 1-D nature more seriously than producing the curvature. Incidently, using the exact form of $\theta$ as in Eq.\ref{eq:theta-lin-fr} as opposed to its approximate form of $C/kx$ does not give the curvature either. Here, we believe that our approach of varying only the directions of ${\bf k'}$s and not their magnitude is a bit oversimplified though it captures the general structure of both the 1-D and 2-D patterns. However, to get both the direction and magnitude of ${\bf k'}$s requires a detailed understanding of the strain field in the affected region. This is not possible from the STM images alone as we have to know the stresses on the boundaries to solve the complete boundary value problem using elasticity theory.

As pointed out earlier, we should be cautious in using this model as this does not describe the quantitative contrast of an STM image. For instance, this model cannot explain the variation in corrugation amplitude of the 2-D super-lattice with its periodicity as seen from Fig.\ref{fig:fft}e. The STM contrast here actually represents a DOS variation as our local tunneling spectra show that the bright regions are more metallic than the dark ones consistent with earlier work by Kuwabara et. al. \cite{kuwabara}. As of now no calculations on electronic DOS on moir\'{e} patterns in graphite exist. A computational model was proposed by Hentschke et. al. \cite{hentschke} and reviewed by Pong et. al. \cite{pong-theory}, which is based on the variation in local density of atoms. In a sense our $I(x,y)$ also quantifies the local density of atoms but our model is more appropriate for the analytical understanding of the large scale structure of the spatially varying moir\'{e} patterns.

\section{Conclusions}
In conclusion, we have studied a spatially varying 2-D and a connected 1-D super-lattice structure on HOPG. This is the first observation of such 1-D fringes connected to a 2-D moir\'{e} lattice and we attribute this pattern to a spatially varying moir\'{e} rotation of a top graphite layer. The spatially varying rotation implies a shear strain in this layer. We have also described a simple model for understanding spatially varying moire\'{e} patterns. This model can successfully simulate the observed pattern by using a spatially varying rotation angle. The 1-D fringes are found to arise from a particular spatial dependence of the moire\'{e} rotation angle.

\section{Acknowledgements}
The financial support from IITK under the `Initiation Grant' scheme and from the MHRD of the Government of India is thankfully acknowledged. S K Choudhary acknowledge financial support from the University Grant Commission, Government of India.


\begin{thebibliography}
\bibitem{cantilever} Simple elasticity theory for a cantilever gives, $tan(\theta)=\frac{dy}{dx}=\frac{6F}{Ybd^3}(L^2-x^2) \approx\theta$ (for small $\theta$) with $F$ as the force acting at one end of the cantilever, $y$ as the vertical deflection at a given $x$. $Y$, $L$, $d$, and $b$ are the Young's modulus, length, thickness, and the width of the cantilever, respectively.
\end{thebibliography}
\end{document}